\documentclass{PoS}

\usepackage{amsfonts} 
\usepackage{amssymb} 
\usepackage{amsmath} 
\usepackage{graphicx}    
\usepackage{subfigure}

\usepackage{multirow}

\title{Isovector and flavor diagonal charges of the nucleon from 2+1+1 flavor QCD}

\ShortTitle{Charges of the nucleon from 2+1+1 flavor QCD}

\author{\speaker{Rajan Gupta}, Tanmoy Bhattacharya, Vincenzo Cirigliano, Boram Yoon  \\
        Theoretical Division T-2, Los Alamos National Laboratory, Los Alamos, NM 87545, USA\\
        E-mail: \email{rg@lanl.gov, cirigliano@lanl.gov, tanmoy@lanl.gov, boram@lanl.gov}}

\author{Yong-Chull Jang\\
        Phsics Department, Brookhaven National Laboratory, Upton, NY 11973, USA\\
        E-mail: \email{ypj@bnl.gov}}

\author{ Huey-Wen Lin \\
  Department of Physics and Astronomy \& Computational Mathematics, Science and Engineering
  Michigan State University, East Lansing, MI 48824
  E-mail: \email{hwlin@pa.msu.edu}}

\author{PNDME Collaboration}

\abstract{We present high-statistics results for the isovector and
  flavor diagonal charges of the proton using 11 ensembles of 2+1+1
  flavor HISQ fermions. In the isospin symmetric limit, results for
  the neutron are given by the $u \leftrightarrow d$ interchange. A
  chiral-continuum fit with leading order corrections was made to
  extract the connected and disconnected contributions in the
  continuum limit and at $M_\pi=135$~MeV. All results are given in the
  $\overline{MS}$ scheme at 2~GeV.  The isovector charges, $g_A^{u-d}
  = 1.218(25)(30)$, $g_S^{u-d} = 1.022(80)(60) $ and $g_T^{u-d} =
  0.989(32)(10)$, are used to obtain low-energy constraints on novel
  scalar and tensor interactions, $\epsilon_{S}$ and $\epsilon_{T}$,
  at the TeV scale.  The flavor diagonal axial charges are: $g_A^u
  \equiv \Delta u \equiv \langle 1 \rangle_{\Delta u^+} =
  0.777(25)(30)$, $g_A^d \equiv \Delta d \equiv \langle 1
  \rangle_{\Delta d^+} = -0.438(18)(30)$, and $g_A^s \equiv \Delta s
  \equiv \langle 1 \rangle_{\Delta s^+} = -0.053(8)$. Their sum gives
  the total quark contribution to the proton spin, $\sum_{q=u,d,s}
  (\frac{1}{2} \Delta q) = 0.143(31)(36)$. This result is in good
  agreement with the recent COMPASS analysis $0.13 < \frac{1}{2}
  \Delta \Sigma < 0.18$.  Implications of results for the flavor
  diagonal tensor charges, $g_T^u = 0.784(28)(10)$, $g_T^d =
  -0.204(11)(10)$ and $g_T^s = -0.0027(16)$ for constraining the quark
  electric dipole moments and their contributions to the neutron
  electric dipole moment are discussed. These flavor diagonal charges also 
  give the strength of the interaction of dark matter with nucleons via 
  axial and tensor mediators. }

\FullConference{The 36th Annual International Symposium on Lattice Field Theory - LATTICE2018\\
		22-28 July, 2018\\
		Michigan State University, East Lansing, Michigan, USA.}

\begin{document}

\section{Introduction}

Lattice QCD provides first principal results for the matrix elements
(ME) of quark bilinear operators within nucleon states that are needed
to quantify a number of properties and structure of the nucleon. All
the calculations reported on here have been done using clover valence
fermions on 2+1+1 flavor HISQ ensembles generated by the MILC
collaboration~\cite{Bazavov:2012xda}.  Detailed analyses and results,
summarized here, for the isovector charges are given in
Ref.~\cite{Gupta:2018qil}, for flavor diagonal axial charges in
Ref.~\cite{Lin:2018obj}, and for flavor diagonal tensor charges in
Ref.~\cite{Gupta:2018lvp}.

High-statistics data were generated cost-effectively using the
truncated solver with bias correction method
(TSM)~\cite{Bali:2009hu}.  We also used the coherent
source method to construct sequential propagators with the insertion
of a zero-momentum nucleon state at the sink time
slice~\cite{Yoon:2016dij}.

\begin{figure*}[hptb]
\begin{center}                                               
  \subfigure{
    \includegraphics[height=1.10in,trim={0.20cm 0.70cm 0 0.06cm},clip]{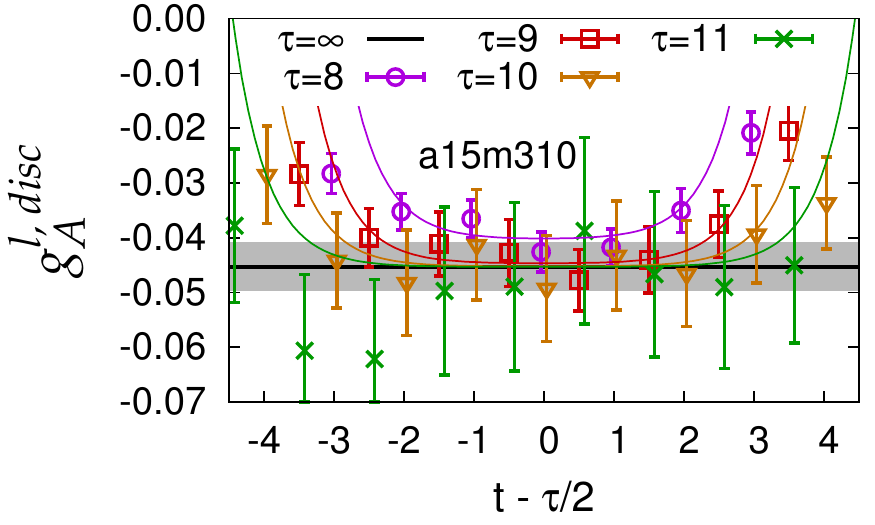}
    \includegraphics[height=1.10in,trim={1.20cm 0.70cm 0 0.06cm},clip]{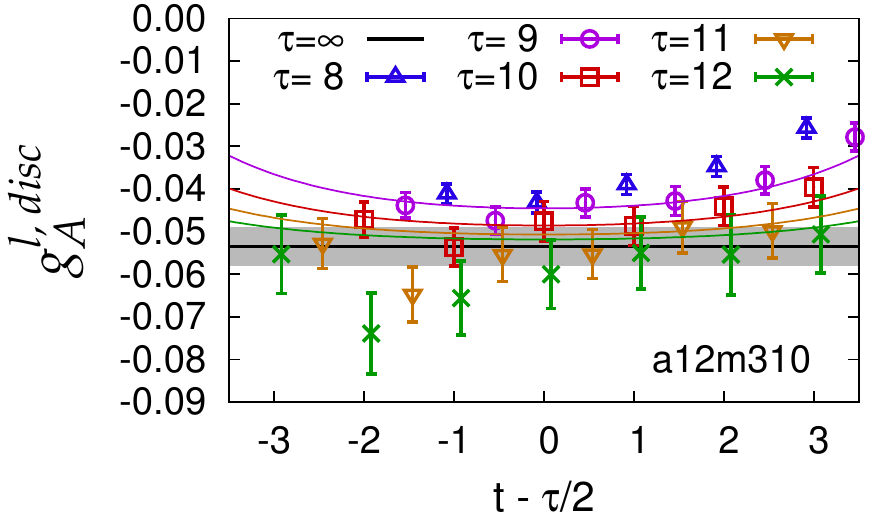}
    \includegraphics[height=1.10in,trim={1.20cm 0.70cm 0 0.06cm},clip]{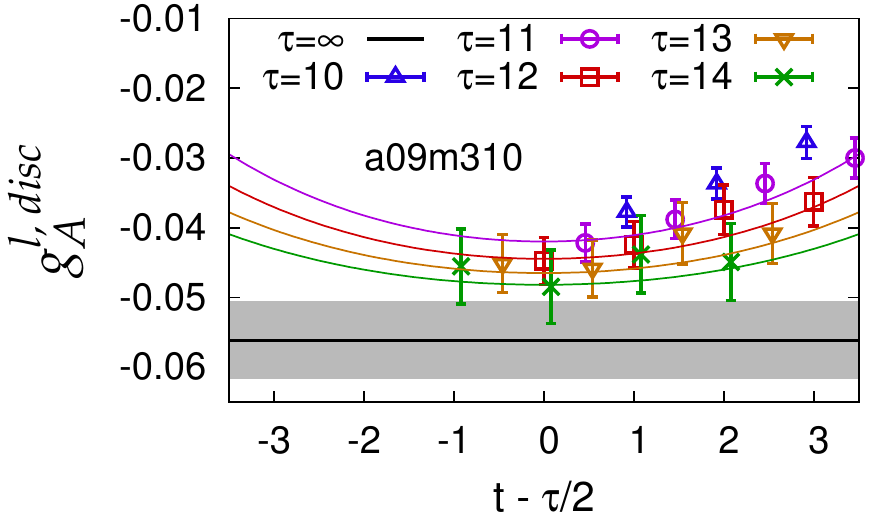}
  }\\
  \subfigure{
    \includegraphics[height=1.24in,trim={0.20cm 0.10cm 0 0.06cm},clip]{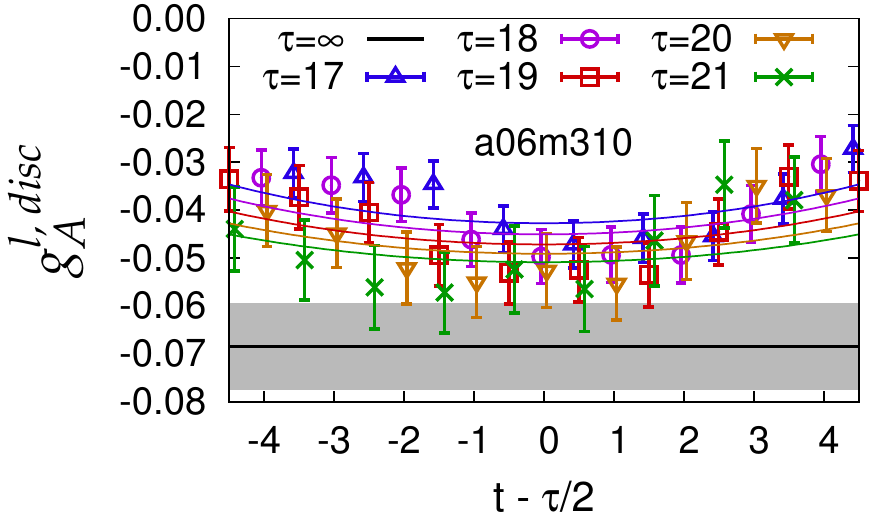}
    \includegraphics[height=1.24in,trim={1.20cm 0.10cm 0 0.06cm},clip]{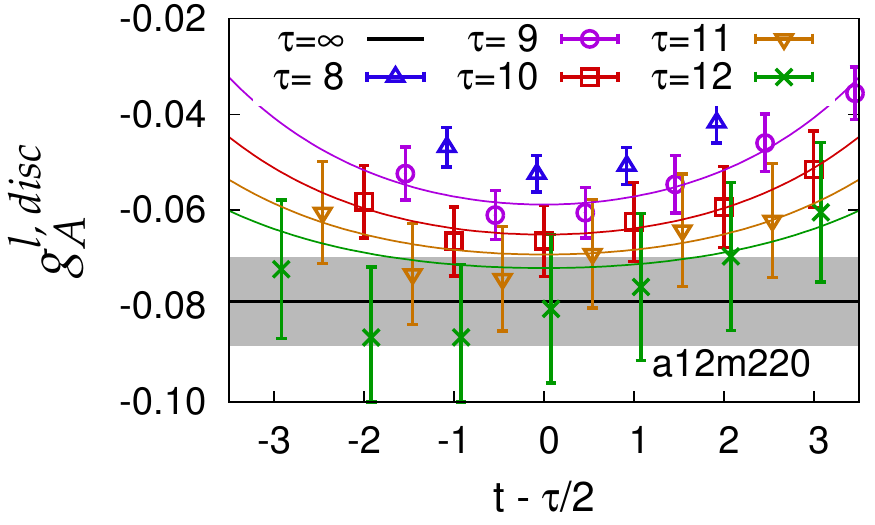}
    \includegraphics[height=1.24in,trim={1.20cm 0.10cm 0 0.06cm},clip]{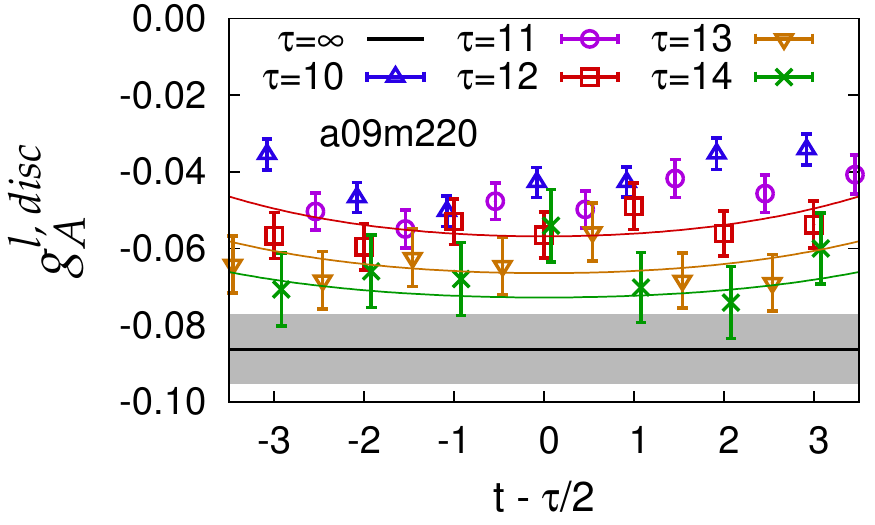}
  }\\
  \subfigure{
    \includegraphics[height=1.10in,trim={0.20cm 0.70cm 0 0.06cm},clip]{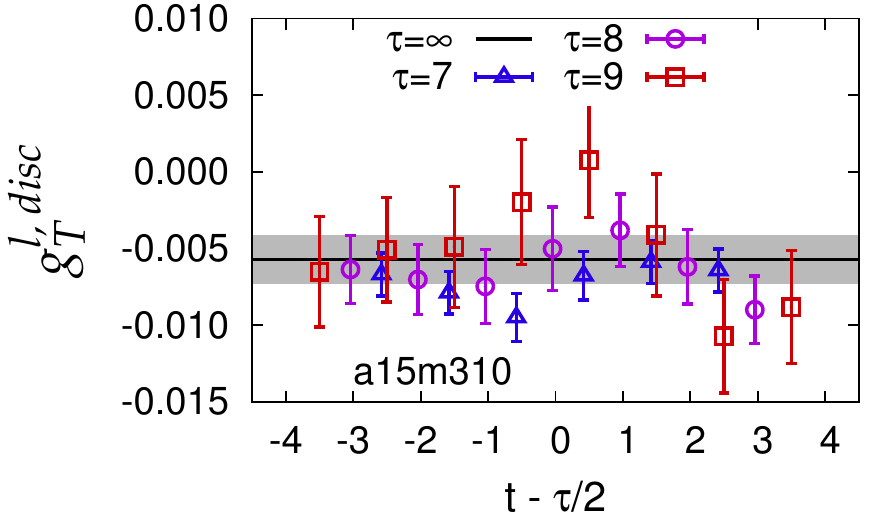}
    \includegraphics[height=1.10in,trim={1.20cm 0.70cm 0 0.06cm},clip]{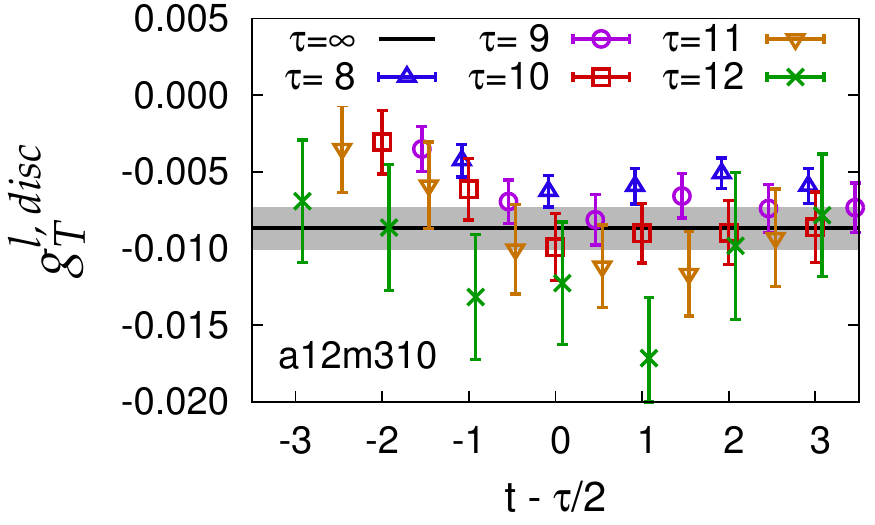}
    \includegraphics[height=1.10in,trim={1.20cm 0.70cm 0 0.06cm},clip]{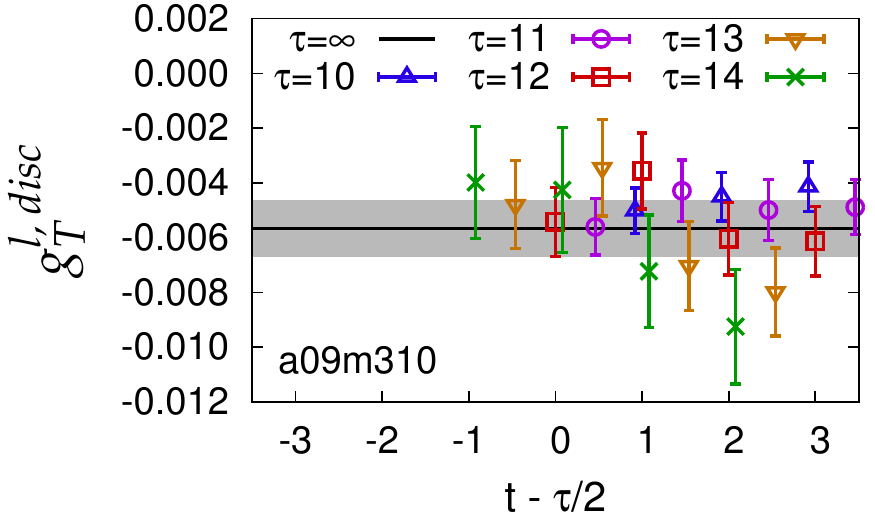}
  }\\
  \subfigure{
    \includegraphics[height=1.24in,trim={0.20cm 0.10cm 0 0.06cm},clip]{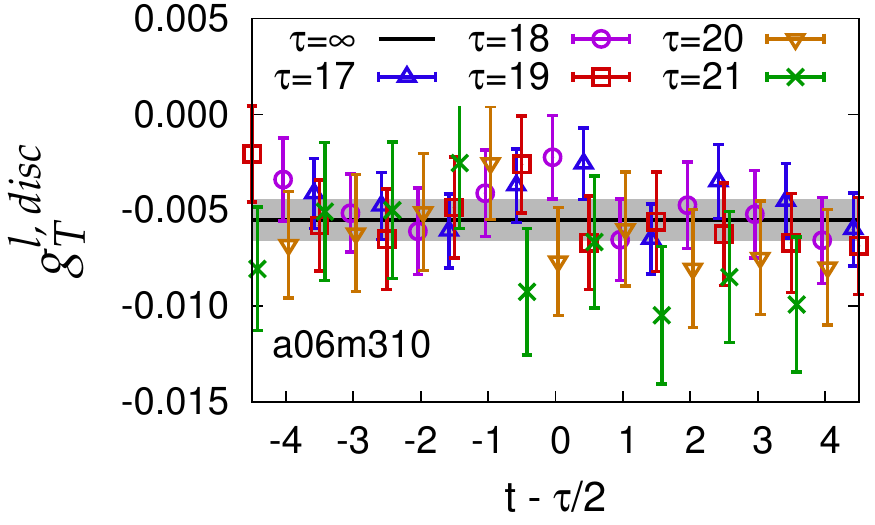}
    \includegraphics[height=1.24in,trim={1.20cm 0.10cm 0 0.06cm},clip]{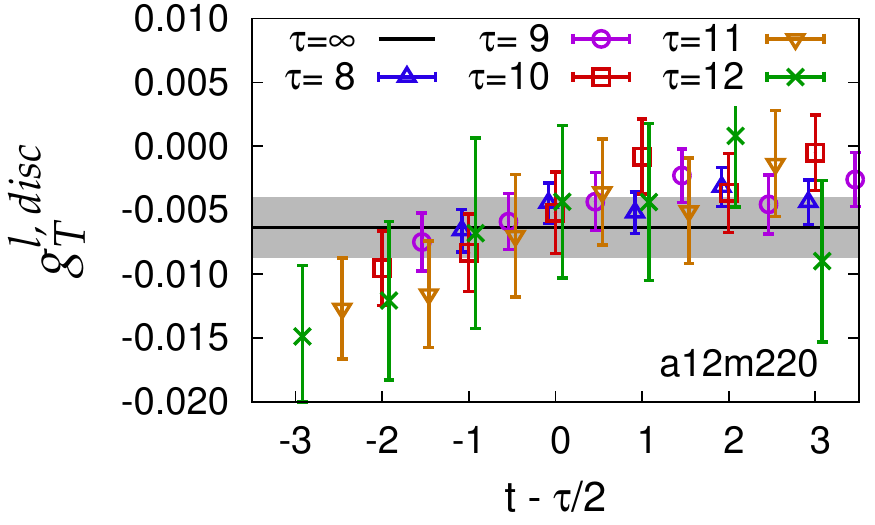}
    \includegraphics[height=1.24in,trim={1.20cm 0.10cm 0 0.06cm},clip]{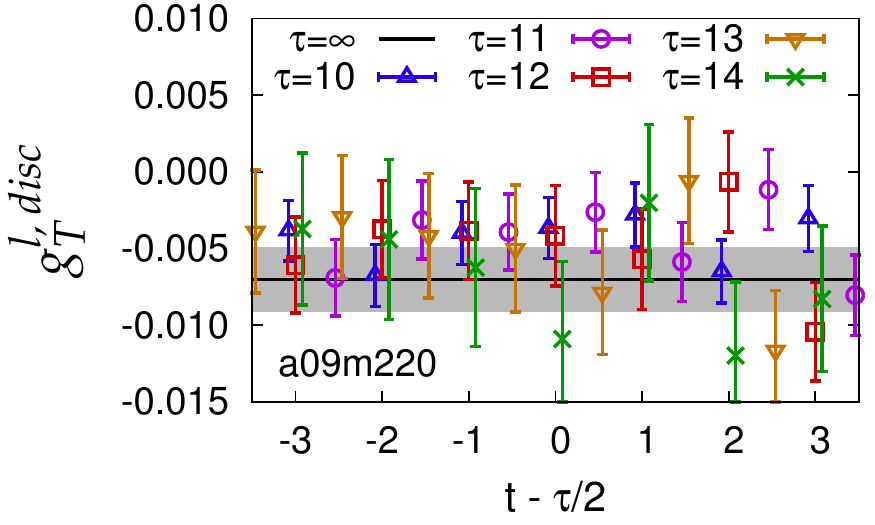}
  }
  \end{center}  
\vspace{-0.4cm}                                                                            
\caption{The data and excited-state fits for the light quark
  disconnected contribution to the bare $g_A^{l,s {\text disc}}$ (top
  2 rows) and $g_A^{l,s {\text disc}}$ (bottom 2 rows) The grey error
  band and the solid line within it is the $\tau \to \infty$ estimate
  obtained using the 2-state (constant) fit to $g_A^{l,\, {\text disc}}$
  ($g_T^{l,\, {\text disc}}$) data at different $t$ and $\tau$. The result
  of the fit for each individual $\tau$ is shown by a solid line in
  the same color as the data points.
  \label{fig:gAgTdisc}}
\end{figure*}

The excited-state contamination (ESC) in the 3-point correlation
functions was analyzed by generating, on each ensemble, data at 3-6
values of the source-sink separation $\tau$. Fits to extract the
ground state ME were then made keeping up to three states in the
spectral decomposition to data at a large set of values of the
operator insertion time $t$ and $\tau$.  The choice of the number of
states kept (1, 2 or 3) depended on the statistical precision of the data and the
size of the ESC. The amplitudes and masses of the ground and excited
states used in these fits were obtained from 2-point functions keeping
four states in fits using the spectral decomposition. The 2-state (one-state, i.e., constant) fits 
to extract the disconnected $g_A^{l}$ ($g_T^{l}$) are shown in Fig.~\ref{fig:gAgTdisc}. 

All the isovector renormalization constants $Z_\Gamma$ have been
determined non-pertubatively in the RI-sMOM scheme and then converted
to $\overline{MS}$ at 2~GeV using 2-loop perturbation theory.  For the
flavor diagonal axial and tensor charges, we assumed that the
difference between $Z^{\rm isovector}$ and $Z^{\rm isoscalar}$, which
is zero at 2-loops~\cite{Constantinou:2016ieh}, is small and
negligible non-perturbatively.

\begin{figure*}[tb]   
\subfigure{
    \includegraphics[width=0.32\linewidth]{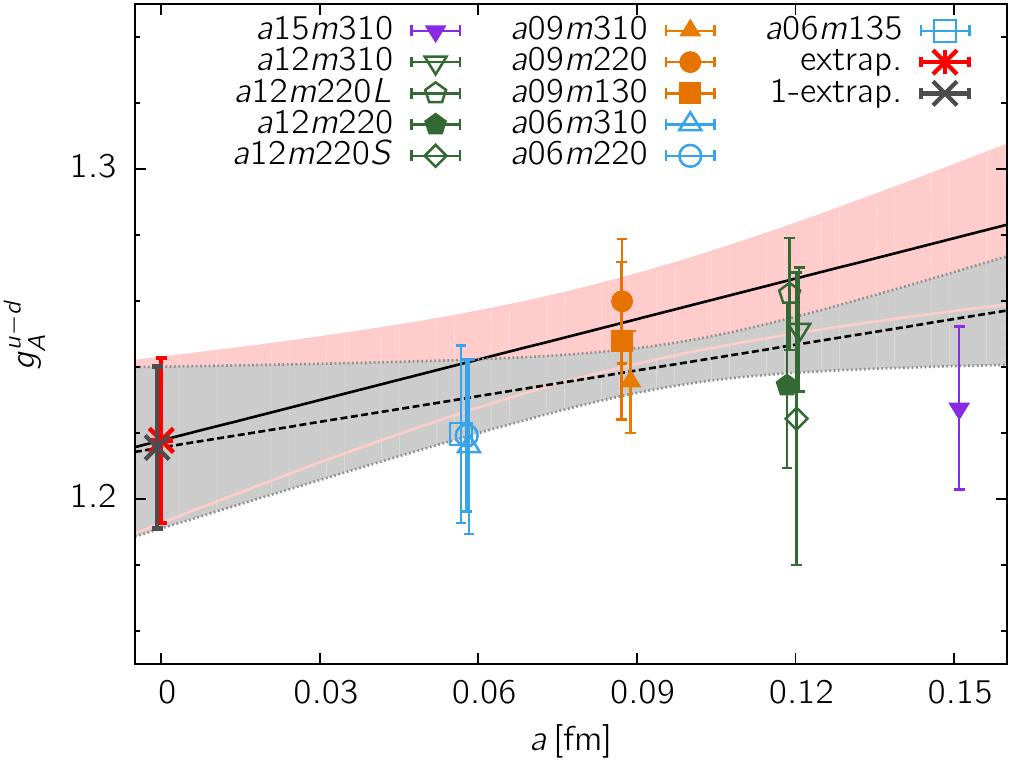}
    \includegraphics[width=0.32\linewidth]{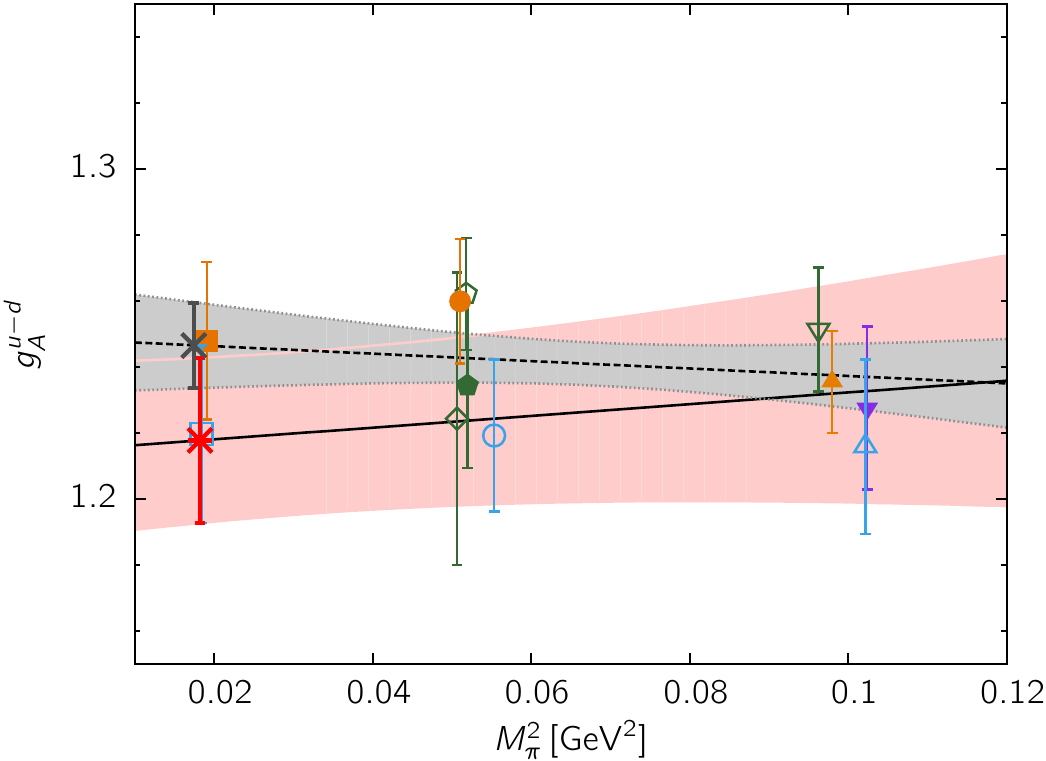}
    \includegraphics[width=0.32\linewidth]{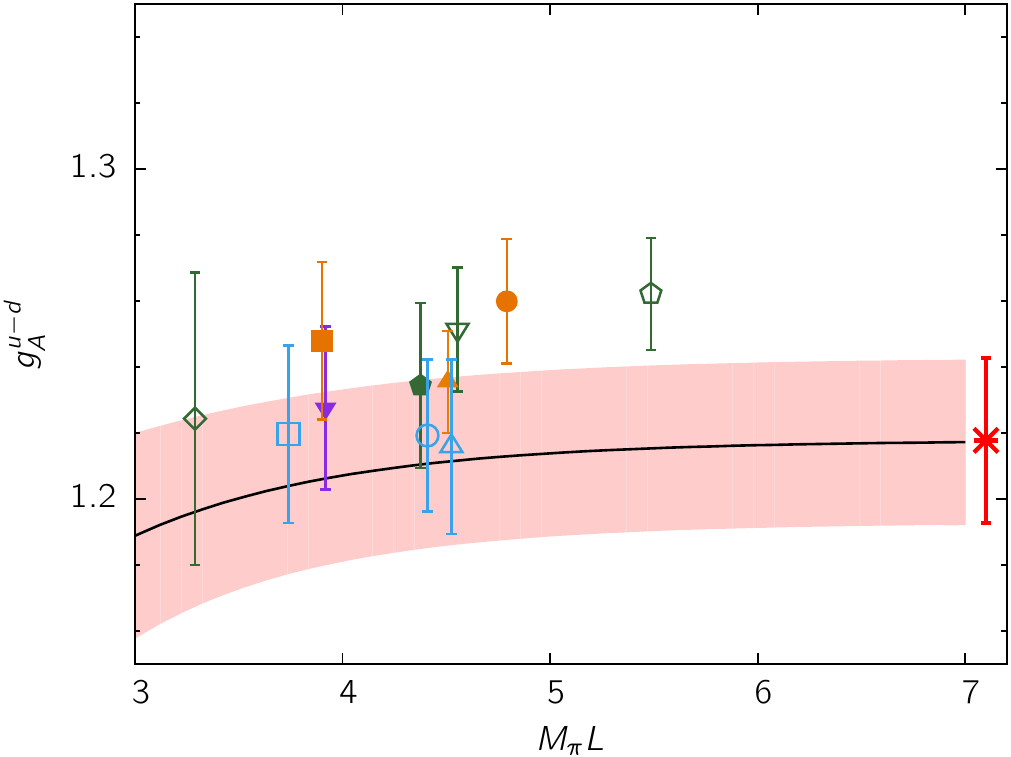}
}
\subfigure{
    \includegraphics[width=0.32\linewidth]{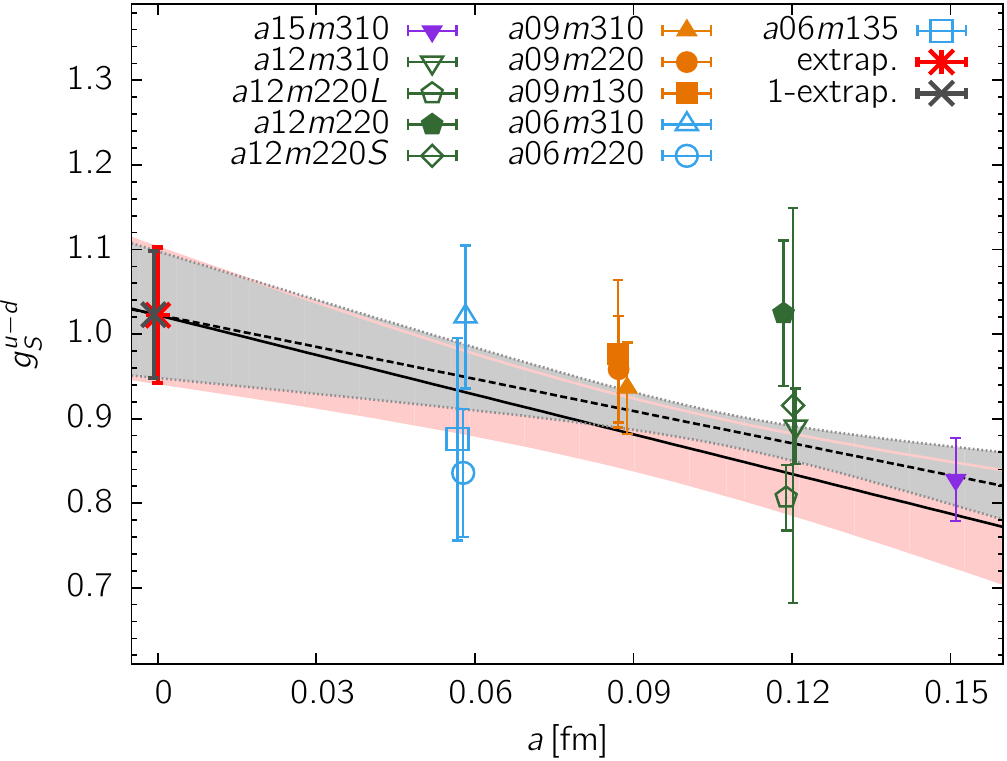}
    \includegraphics[width=0.32\linewidth]{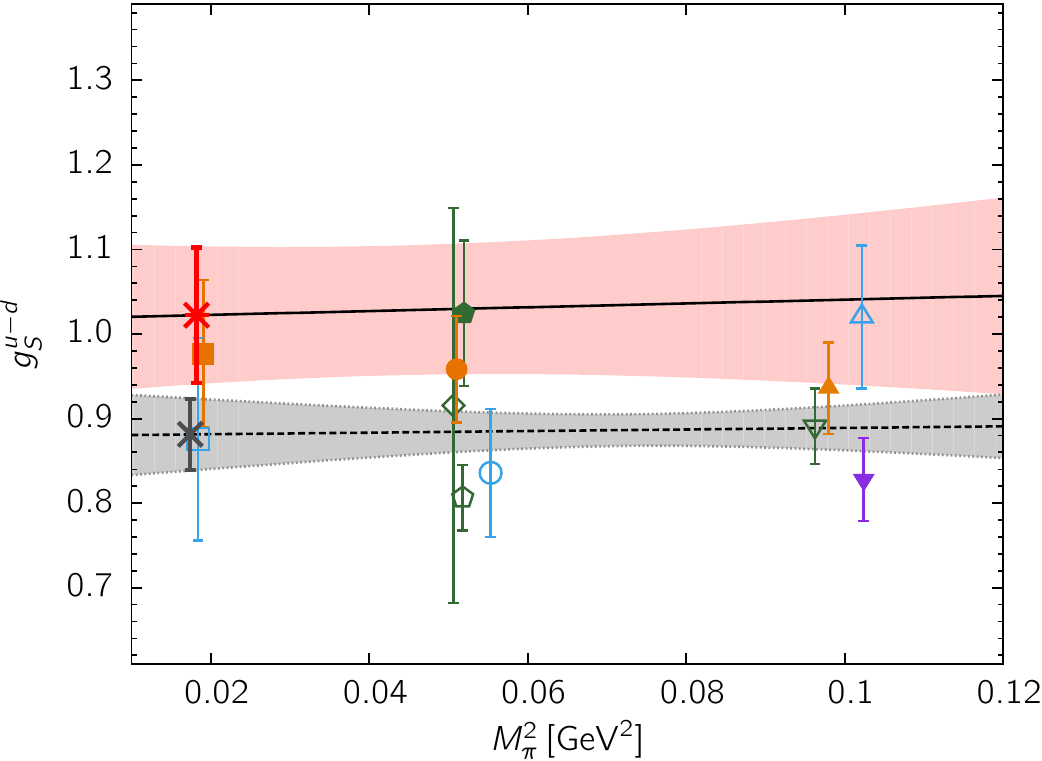}
    \includegraphics[width=0.32\linewidth]{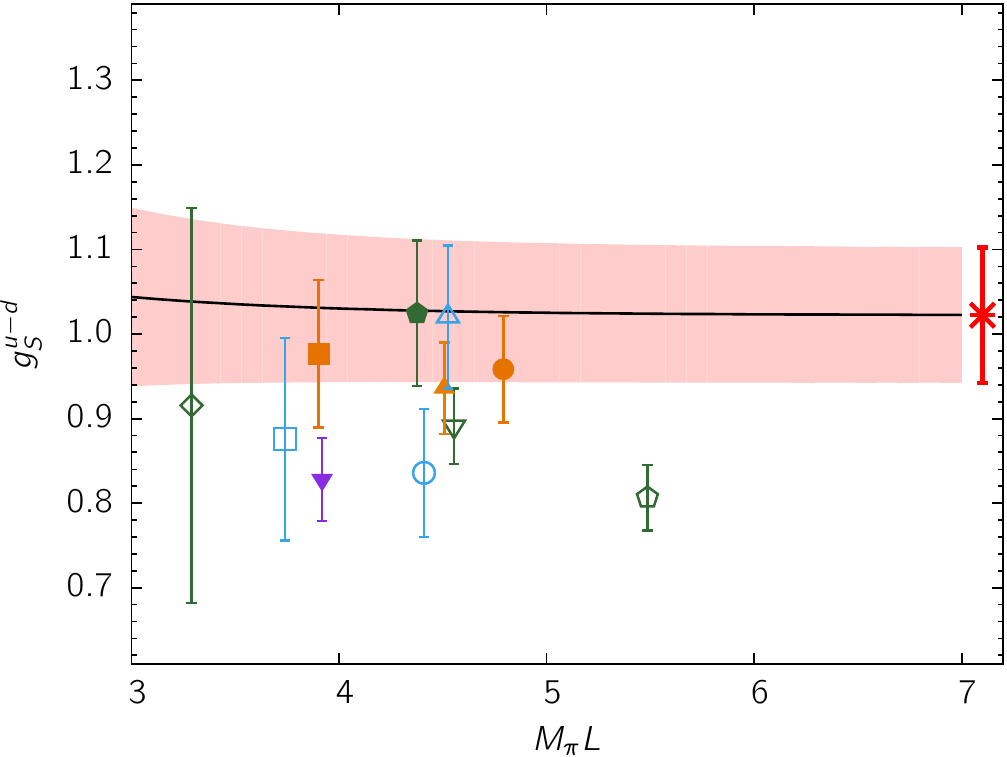}
}
\subfigure{
    \includegraphics[width=0.32\linewidth]{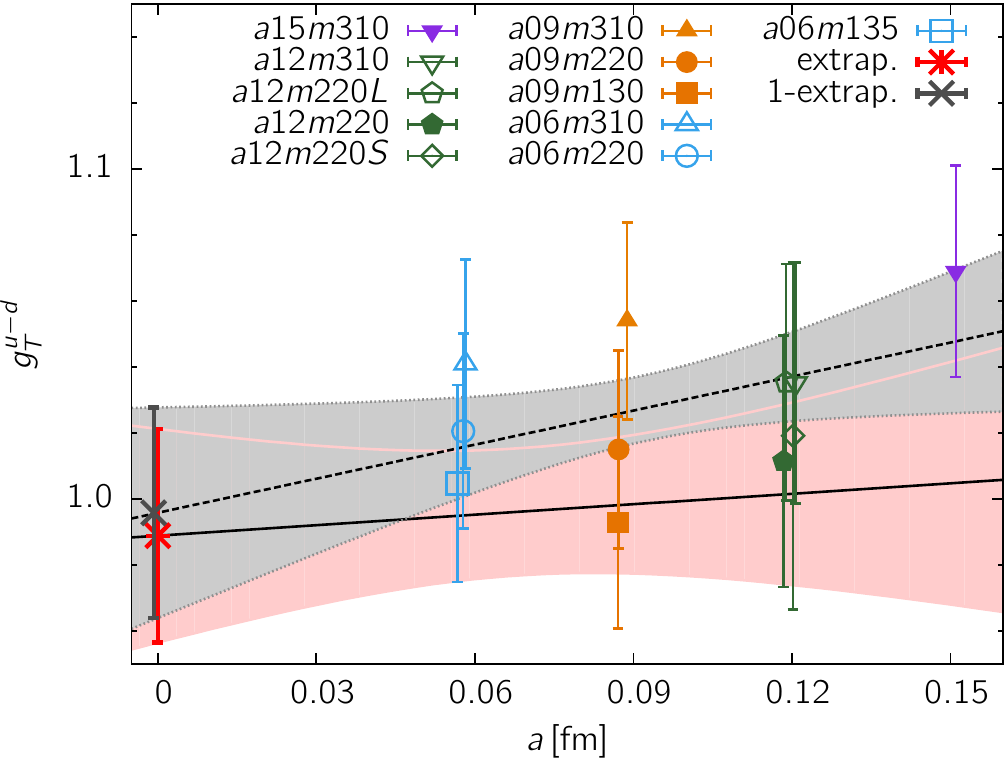}
    \includegraphics[width=0.32\linewidth]{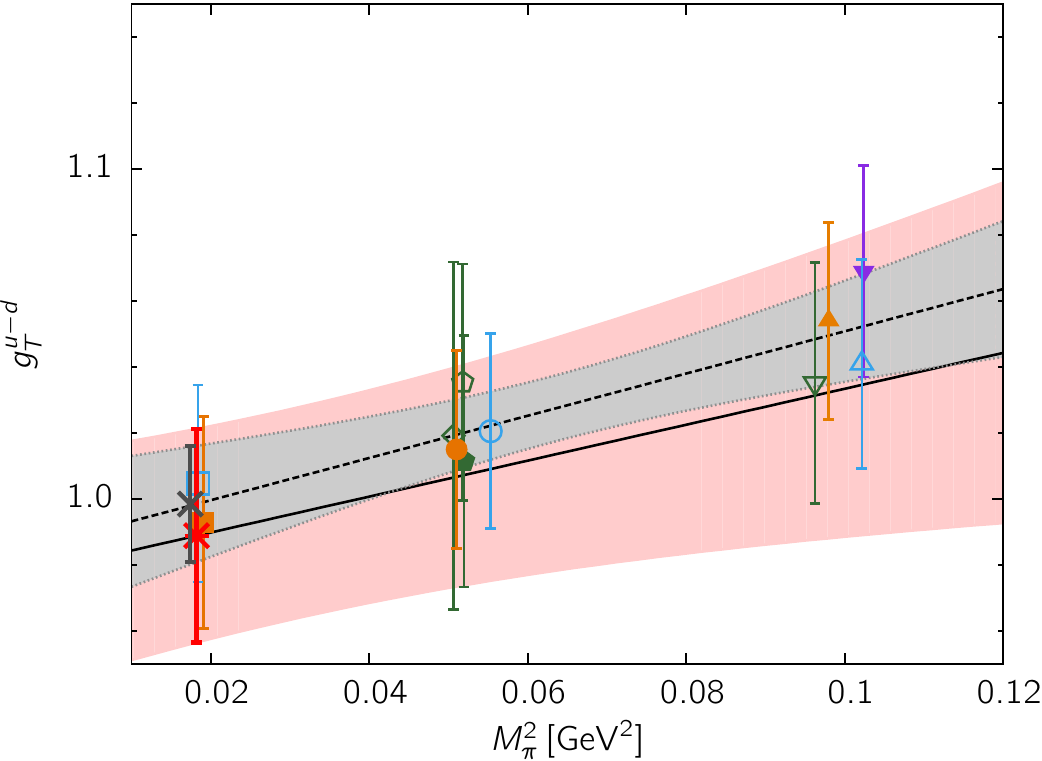}
    \includegraphics[width=0.32\linewidth]{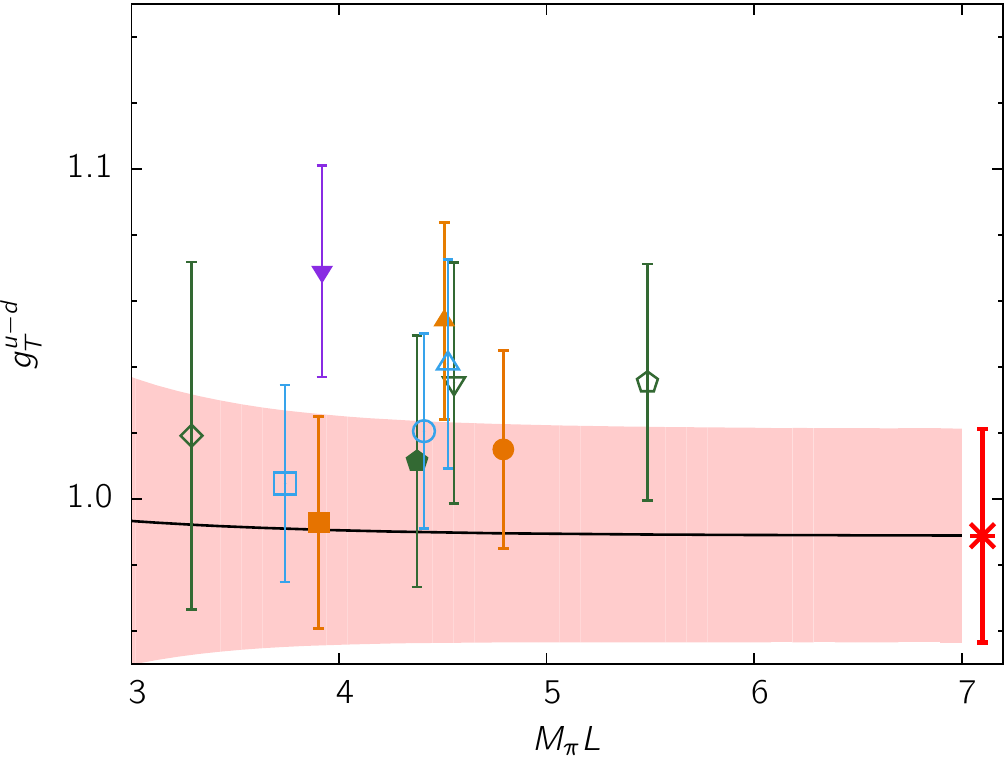}
}
\caption{The 11-point CCFV
  fit using Eq.~\protect\eqref{eq:extrapgAST} to the data for the
  renormalized isovector charges $g_A^{u-d}$, $g_S^{u-d}$, and
  $g_T^{u-d}$ in the $\overline{{\rm MS}}$ scheme at 2~GeV. The
  result of the simultaneous extrapolation to the physical point
  at $a\rightarrow 0$, $M_\pi \rightarrow M_{\pi^0}^{{\rm
      phys}}=135$~MeV and $M_\pi L \rightarrow \infty$ are marked by a red
  star.  The pink error band in each panel is the result of the
  simultaneous fit but shown as a function of a single variable. The
  overlay in the left (middle) panels with the dashed line within the
  grey band is the fit to the data versus $a$ ($M_\pi^2$), i.e.,
  neglecting dependence on the other two variables. The symbols used to plot the data are 
  defined in the left panels.}
  \label{fig:conUmD-extrap11}
\end{figure*}

\begin{figure*}[th]
\centering
  \subfigure{
    \includegraphics[height=1.02in,trim={0.1cm   0.10cm 0 0.1cm},clip]{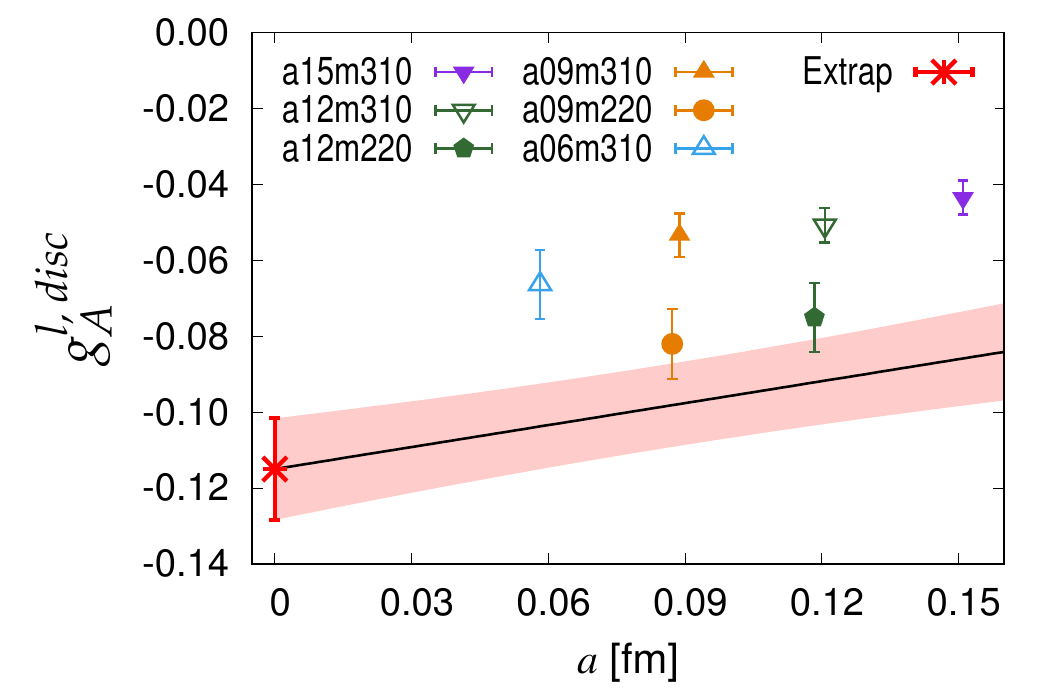} 
    \includegraphics[height=1.02in,trim={1.3cm   0.10cm 0 0.1cm},clip]{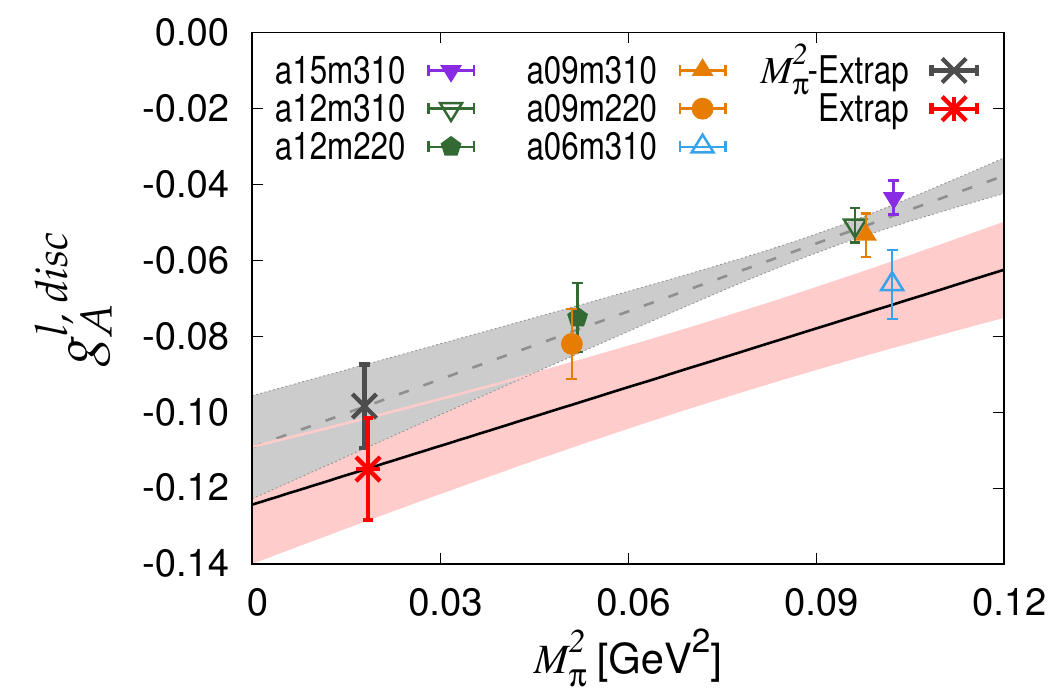}
    \includegraphics[height=1.02in,trim={0.1cm   0.10cm 0 0.1cm},clip]{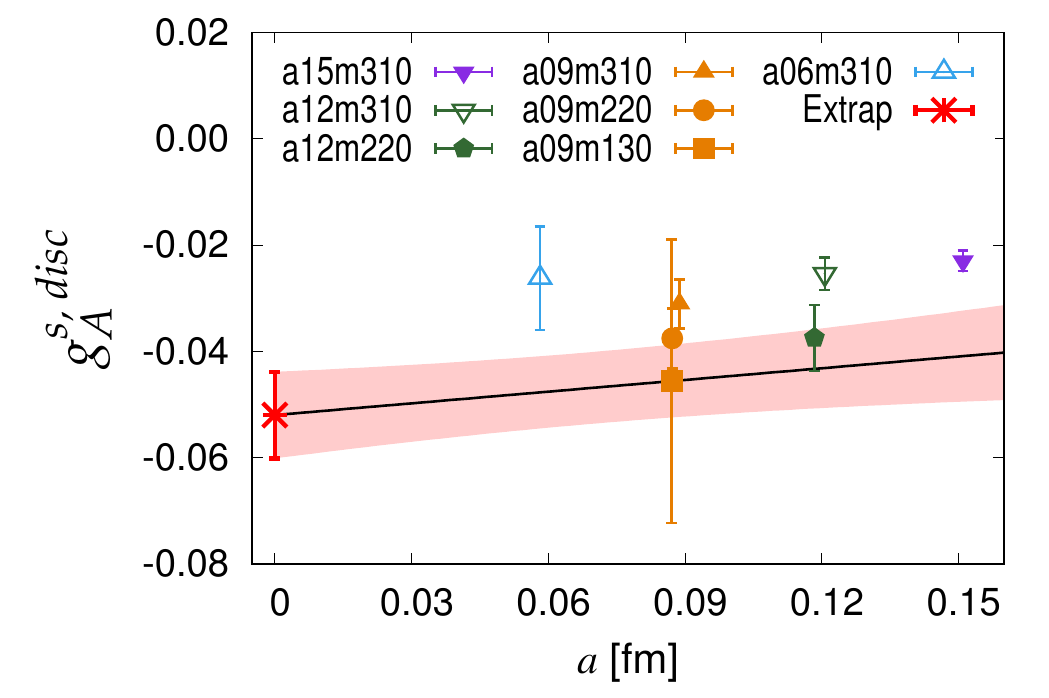} 
    \includegraphics[height=1.02in,trim={1.3cm   0.10cm 0 0.1cm},clip]{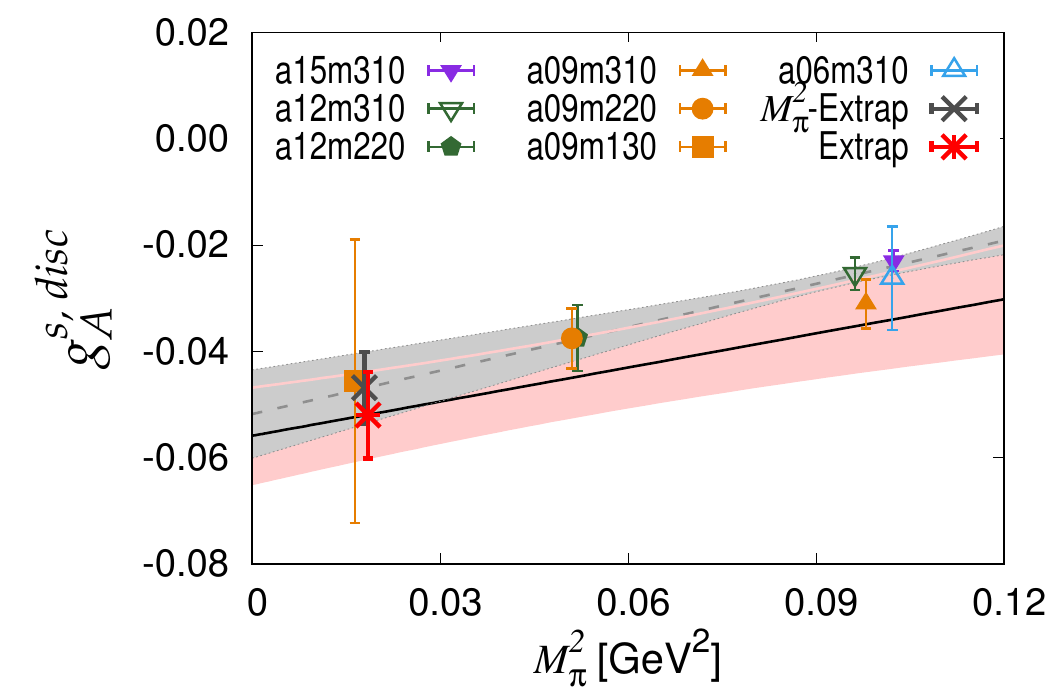} 
  }
  \subfigure{
    \includegraphics[height=1.02in,trim={0.1cm   0.10cm 0 0.1cm},clip]{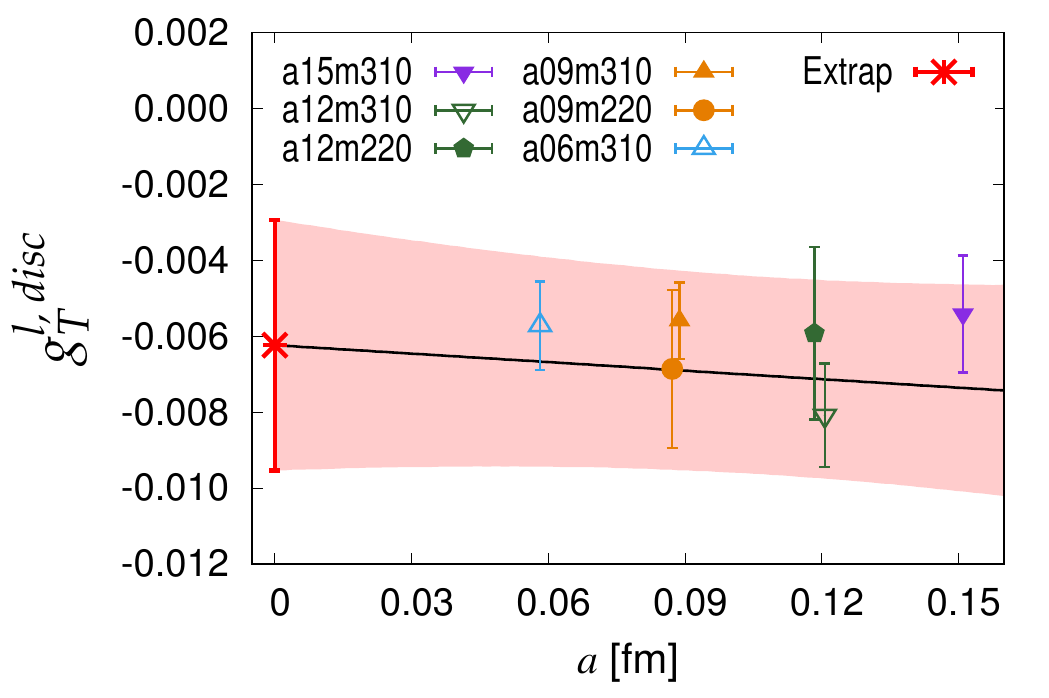} 
    \includegraphics[height=1.02in,trim={1.3cm   0.10cm 0 0.1cm},clip]{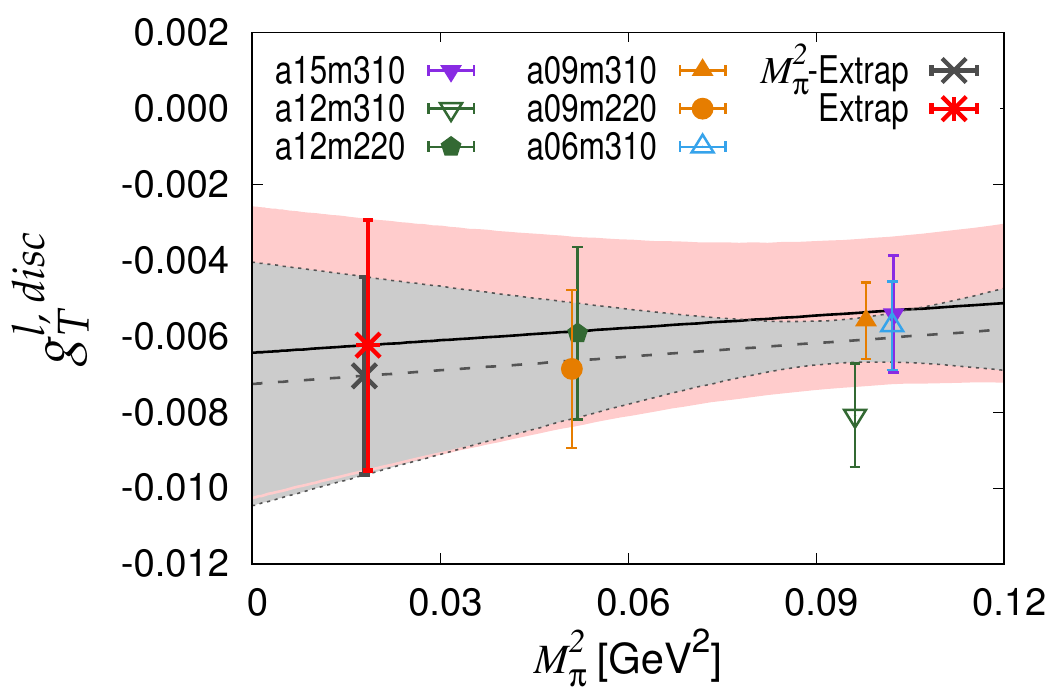}
    \includegraphics[height=1.02in,trim={0.1cm   0.10cm 0 0.1cm},clip]{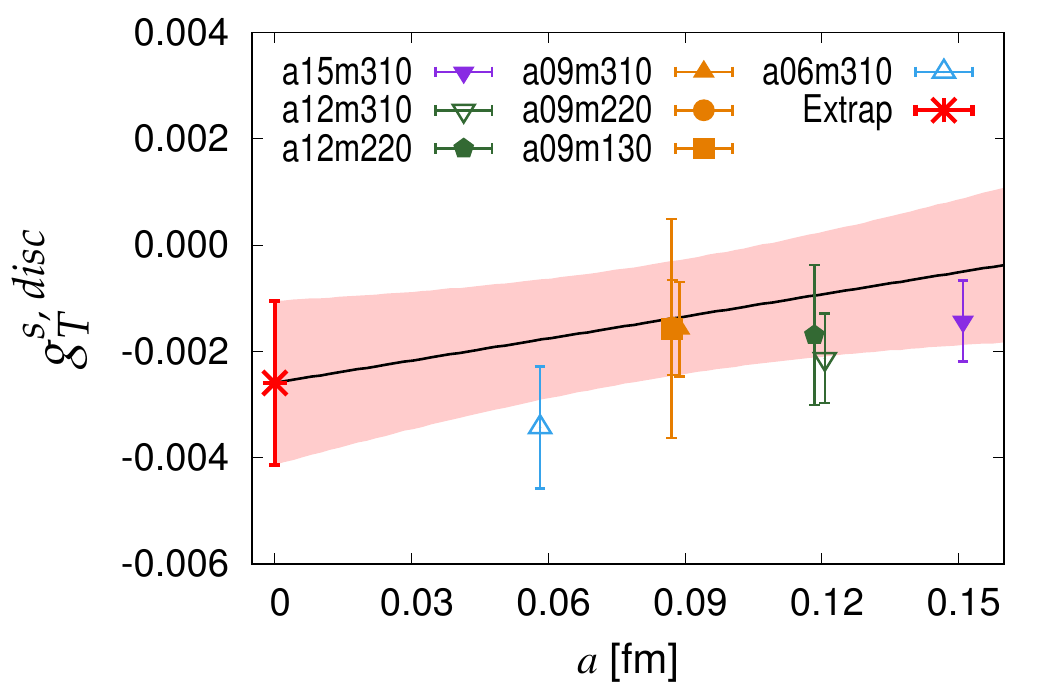} 
    \includegraphics[height=1.02in,trim={1.3cm   0.10cm 0 0.1cm},clip]{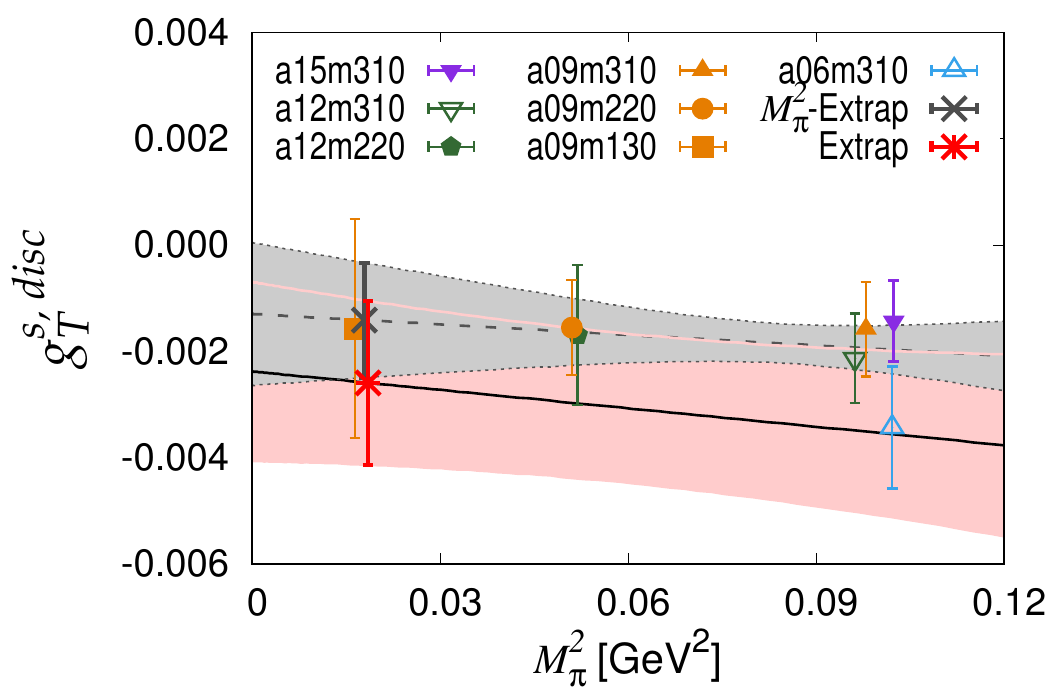} 
  }
\vspace{-0.3cm}
\caption{The extrapolation of the renormalized disconnected $g_A^{l,s}$ (top row)
  and $g_T^{l,s}$ data using the chiral-continuum ansatz
  given in Eq.~\protect\eqref{eq:extrapgAST}.  In each panel, the pink
  band shows the result of the simultaneous fit plotted versus a
  single variable with the other variable set to its physical
  value. The result at the physical point, $M_\pi = 135$~MeV and
  $a=0$, is marked with a red star. The grey band
  shows the fit versus only $M_\pi$, i.e., ignoring the dependence on
  $a$. It highlights the need for a simultaneous fit in both
  $a$ and $M_\pi$.  \looseness-1
  \label{fig:gls-extrap}}
\end{figure*}

To obtain estimates of the renormalized charges in the continuum limit
($a\rightarrow 0$), physical pion mass ($M_{\pi^0} = 135$~MeV) and in
the infinite volume limit ($L \rightarrow \infty$), we used a
simultaneous chiral-continuum-finite-volume (CCFV) fit ansatz with 
leading order corrections:
\begin{equation}
g_{A,S,T} (a,M_\pi,L) = c_1 + c_2 a + c_3 M_\pi^2 + c_4 M_\pi^2 e^{-M_\pi L} \,. 
\label{eq:extrapgAST}
\end{equation}
For our clover-on-HISC calculation, the leading discretization
correction is $O(a)$ since the clover action and operators are not
fully $O(a)$ improved. The CCFV fits for the isovector charges using
data on 11 ensembles are shown in Fig.~\ref{fig:conUmD-extrap11}.  The
disconnected contributions are small in magnitude, have larger
statistical errors and were analyzed on fewer ensembles, 6 for light
quarks and 7 for the strange. The chiral-continuum extrapolation (neglecting the
finite volume corrections as they were 
small in the connected contributions) for disconnected $g_A^{l,s}$
and $g_T^{l,s}$ is shown in Fig.~\ref{fig:gls-extrap}.
\vspace{-0.15cm}


%


\section{Results}

Using the CCFV fit ansatz given in Eq.~\eqref{eq:extrapgAST}, the
continuum limit results for the isovector charges of the proton, at $M_\pi=135$~MeV and 
in the $\overline{MS}$ scheme at 2~GeV, are
\begin{equation}
g_A^{u-d} = 1.218(25)(30); \qquad g_S^{u-d} = 1.022(80)(60), \qquad g_T^{u-d} = 0.989(32)(10) \,.
\label{eq:givAST}
\end{equation}
The results for the flavor diagonal axial and tensor charges, in the $\overline{MS}$ scheme at 2~GeV,  are:
\begin{equation}
g_A^u \equiv \Delta u \equiv \langle 1 \rangle_{\Delta u^+} = 0.777(25)(30), \, 
g_A^d \equiv \Delta d \equiv \langle 1 \rangle_{\Delta d^+} = -0.438(18)(30), \, 
g_A^s \equiv \Delta s = -0.053(8) \,.
\label{eq:gAfd}
\end{equation}
\begin{equation}
g_T^u = 0.784(28)(10), \qquad g_T^d = -0.204(11)(10), \qquad  g_T^s = -0.0027(16) \,.
\label{eq:gTfd}
\end{equation}
The second error accounts for the uncertainty due to the chiral-continuum ansatz, Eq.~\eqref{eq:extrapgAST} . \looseness-1

\section{Phenomenology}

The isovector axial charge is a benchmark quantity for the
calculations of nucleon matrix elements since it is known with high
accuracy experimentally. Our result, $g_A^{u-d} = 1.218(25)(30)$, is
about $5\%$ or $1.5\sigma$ below the latest experimental value
$g_A^{u-d} = 1.2772(20)$~\cite{Brown:2017mhw} and the recent CalLat
result, $g_A^{u-d} = 1.271(13)$~\cite{Chang:2018uxx}.  In
Ref.~\cite{Gupta:2018qil}, we show that the difference from the CalLat
result comes from differences in the chiral and continuum
extrapolation. CalLat has not simulated the three HISQ ensembles at
$a\approx 0.57$~fm, which pull down our result (see
Fig.~\ref{fig:conUmD-extrap11}). Also, their data on the two physical
mass ensembles have large errors and do not contribute significantly
to the fits, i.e., their estimate is essentially based on fits to data
with $M_\pi \ge 220 $~MeV. Our conclusion is that more extensive
calculations (larger range of $a$, more high statistics physical mass ensembles and
different actions) are needed to understand and quantify the
systematics at the percent level.

\begin{figure*}
\begin{center}
\includegraphics[width=.49\textwidth,clip]{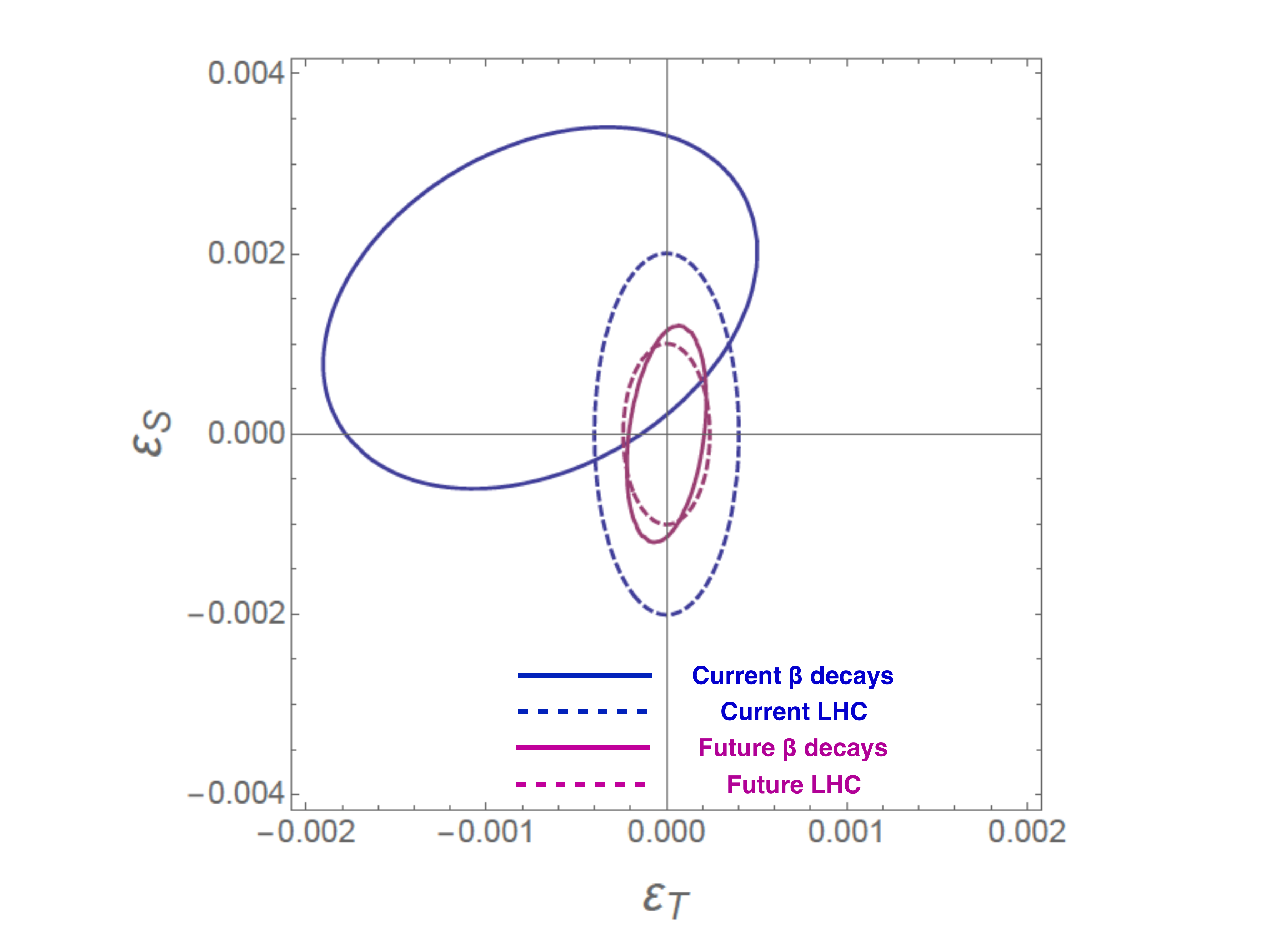}
\includegraphics[width=.49\textwidth,clip]{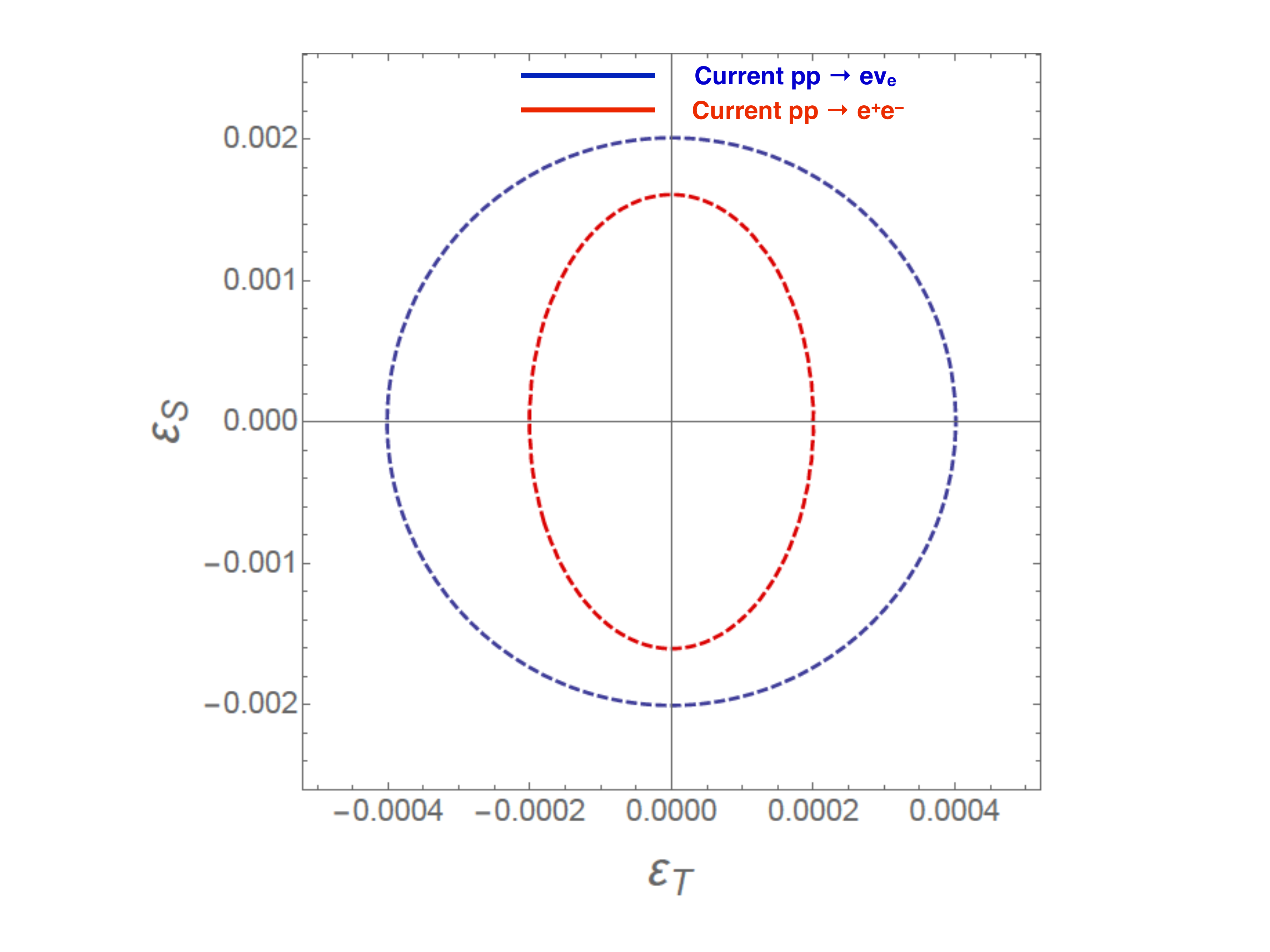}
\end{center}
\vspace{-0.5cm}
\caption{Current and projected $90 \%$ C.L. constraints on
  $\epsilon_S$ and $\epsilon_T$ defined at 2~GeV in the
  $\overline{MS}$ scheme.  (Left) The beta-decay constraints are
  obtained from the recent review article
  Ref.~\protect\cite{Gonzalez-Alonso:2018omy}. The current and future
  LHC bounds are obtained from the analysis of the $pp \to e + MET +
  X$.  We have used the ATLAS results~\protect\cite{Aaboud:2017efa},
  at $\sqrt{s} = 13$~TeV and integrated luminosity of 36 fb$^{-1}$.
  We find that the strongest bound comes from the cumulative
  distribution with a cut on the transverse mass at 2 TeV.  The
  projected future LHC bounds are obtained by assuming that no events
  are observed at transverse mass greater than 3 TeV with an
  integrated luminosity of 300 fb$^{-1}$. (Right) Comparison of
  current LHC bounds from $pp \to e + MET + X$ versus $pp \to e^+ e^-
  + X$.  }
\label{fig:eSeT}
\end{figure*}

The scalar, $g_S^{u-d}$, and tensor, $g_T^{u-d}$, charges, combined
with low energy neutron decay experiments, provide bounds on novel
scalar and tensor interactions that can arise in BSM models at the TeV
scale~\cite{Bhattacharya:2011qm}.  Our results, given in Eq.~\eqref{eq:givAST}, have reached
the 10\% accuracy needed to complement measurements of helicity flip
parameters $b$ and $B$ in neutron decay at the $10^{-3}$
level~\cite{Bhattacharya:2011qm}.  In Fig.~\ref{fig:eSeT}, we compare
present and future bounds from low-energy and LHC experiments, and
find that, to match future LHC bounds, low energy experiments
need to achieve better than $10^{-3}$ accuracy.\looseness-1

The quark spin contribution to the spin of the proton is given 
by $\sum_{q=u,d,s} (\frac{1}{2} \Delta q) = 0.143(31)(36)$ using the 
results given in Eq.~\eqref{eq:gAfd}. This estimate is obtained without
model assumptions and is in good agreement with the recent COMPASS
analysis $0.13 < \frac{1}{2} \Delta \Sigma < 0.18$~\cite{Adolph:2015saz}. 

\begin{figure*}[tbh]
\begin{flushleft}                                               
  \subfigure{
    \includegraphics[width=.55\textwidth]{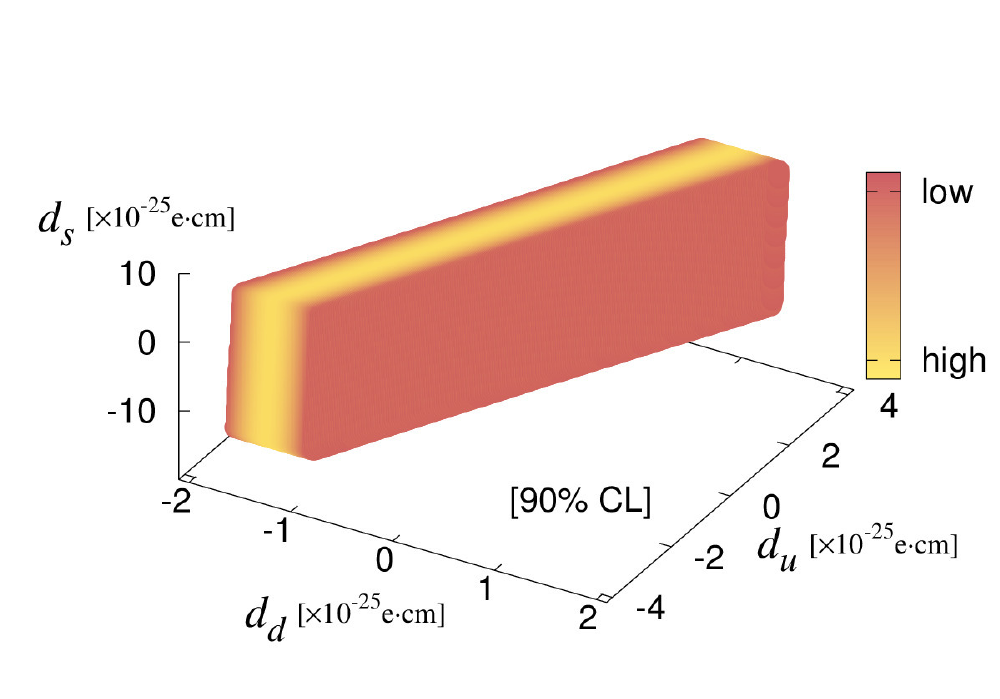} 
    \includegraphics[width=.48\textwidth]{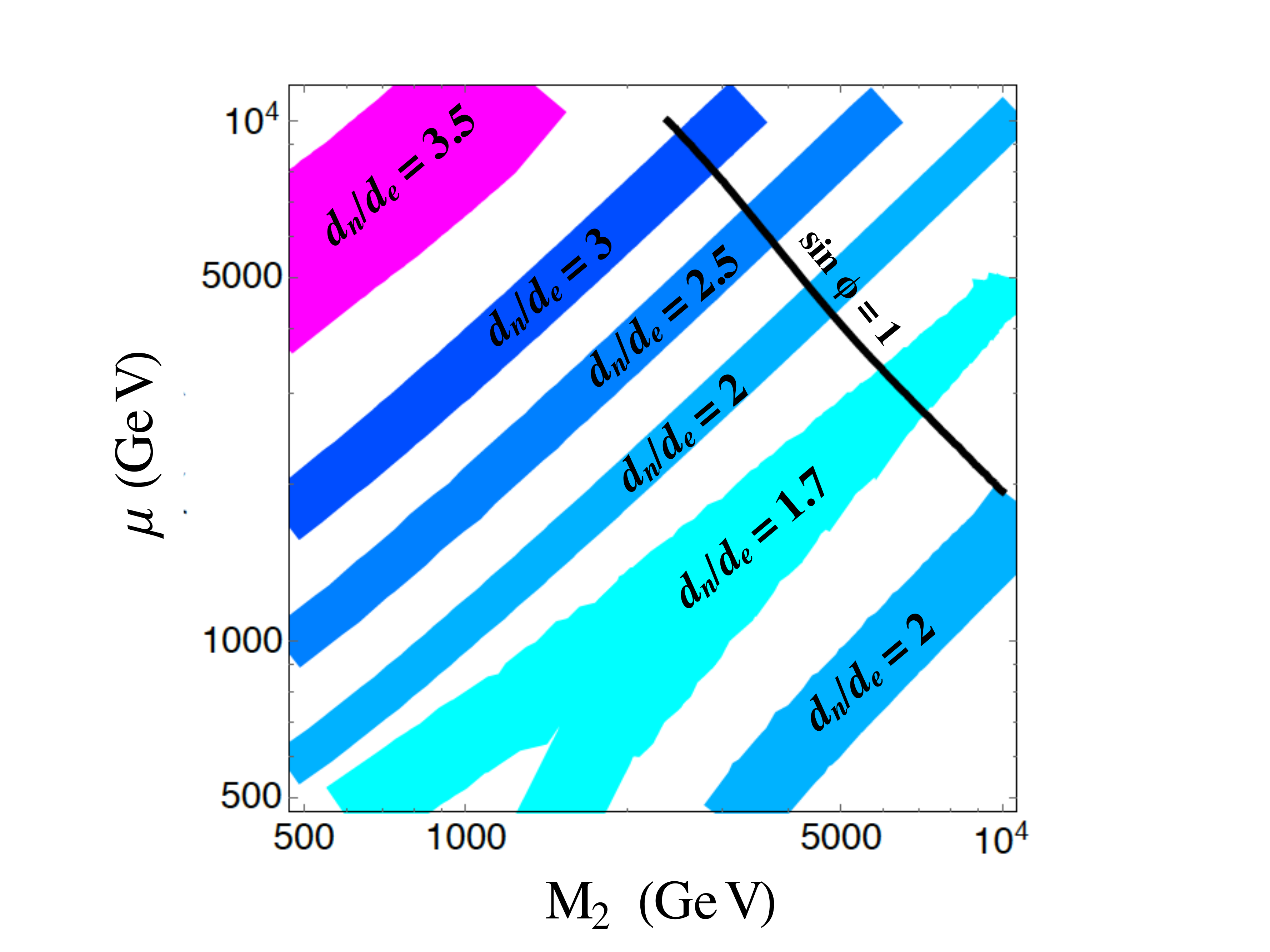} 
  }
 \end{flushleft}                                                      
\caption{(Left) Constraints on the BSM couplings of the CP-violating
  quark EDM operator using the current experimental bound on the nEDM
  ($2.9\times 10^{-26}\ e$ cm~\protect\cite{Baker:2006ts}) and
  assuming that only these couplings contribute. The strongest
  constraint is a strip in $d_u$ and $d_d$, i.e., representing the
  thickness of the slab, with high (low) corresponding to a p-value =
  1 (0.1). (Right) Regions in $M_2$-$\mu$ plane corresponding to
  various values of $d_n/d_e$ in split SUSY, obtained by varying
  $g_T^{u,d,s}$ within our estimated uncertainties.  In the bands of
  constant $d_n/d_e$, the values of both $d_n$ and $d_e$ decrease as
  $\mu$ and $M_2$ increase. Using $d_e \le 1.1 \times 10^{-29} $ e
  cm~\protect\cite{Andreev:2018ayy} and assuming maximal CP violation
  ($\sin \phi = 1$), the allowed region lies above the solid black
  line. For $\mu, M_2 > 500$~GeV, maximizing the ratio $d_n/d_e$ along
  this line gives the upper bound $d_n < 4.1 \times 10^{-29}$ e cm at 
  $d_n/d_e=3.71$.
\label{fig:nEDM}}
\end{figure*}

The contribution of the quark electric dipole moment operator for each quark flavor
to the neutron electric dipole moment is given by 
\begin{equation}
d_n = d_u^\gamma g_T^u + d_u^\gamma g_T^u + d_u^\gamma g_T^u + \ldots \,,
\label{eq:nEDM}
\end{equation}
where $g_T^{u,d,s}$ are the flavor diagonal tensor charges 
given in Eq.~\eqref{eq:gTfd} with $u \leftrightarrow d$ interchange for neutrons. Using these and the
current experimental bound on the nEDM, $2.9\times 10^{-26}\ e$
cm~\protect\cite{Baker:2006ts}, we can constrain the quark EDMs
$d_{u,d,s}^\gamma$ as shown in Fig.~\ref{fig:nEDM} (left). Conversely,
given the values of $d_{u,d,s}^\gamma$ for a given BSM model, one can
use our results for $g_T^{u,d,s}$ to bound $d_n$ provided quark EDMs are the 
dominant CP-violating operators that arise in BSM theories since 
each CP-violating operator contributes to $d_n$ with terms 
analogous to Eq.~\eqref{eq:nEDM}. It turns out that in the ``split
SUSY''
model~\cite{ArkaniHamed:2004fb,Giudice:2004tc,ArkaniHamed:2004yi}, the
fermion EDM operators provide the dominant BSM source of CP
violation~\cite{Bhattacharya:2015esa}.  For this model, the allowed
contour plots for $d_n/d_e$ in the gaugino ($M_2$) and Higgsino
($\mu$) mass parameter plane with the range $500$~GeV to $10$~TeV are
shown in Fig.~\ref{fig:nEDM} (right).  For this analysis, we have
followed Ref.~\cite{Giudice:2005rz} and set $\tan \beta = 1$.  In
particular, our results, assuming maximal CP violation ($\sin \phi =
1$), and the recent experimental bound $d_e < 1.1 \times 10^{-29}
e$~cm~\cite{Andreev:2018ayy}, imply the split-SUSY upper
bound $d_n < 4.1 \times 10^{-29} e$~cm. This limit is falsifiable by
the next generation nEDM experiments.

\acknowledgments
We thank the MILC collaboration for sharing the $2+1+1$-flavor HISQ
ensembles generated by them. We gratefully acknowledge the computing
facilities at and resources provided by NERSC, Oak Ridge OLCF, USQCD
and LANL Institutional Computing.

\bibliographystyle{JHEP}
\bibliography{ref} 

\end{document}